# From Replacement to Orchestration: A Socio-Technical Architecture for Agentic AI in Corporate R&D

## The HARMONY Model

Haithem Boussaid[1•] · Marc Heemskerk[2] · Jimmy Siméon[3] · Adam Breen[4] · Merouane Debbah[5]

[1] EmLyon Business School, France

[2] ASML, Netherlands

[3] AI Medical, France

[4] Independent Consultant, France

[5] Digital Future Institute, Khalifa University, UAE

Correspondence: haithem.boussaid@edu.executive.em-lyon.com

---

## Abstract

**Purpose:** Corporate R&D faces a persistent productivity paradox: rising investment and expanding scientific knowledge have not translated into proportional innovation output. In pharmaceuticals this is captured as Eroom's Law; analogous patterns appear across engineering, materials science, and healthcare. The core cause is not insufficient tools but cognitive saturation: researchers spend an increasing share of their effort on coordination, documentation, and data governance -- hidden work that displaces high-value hypothesis formation, interpretation, and strategic synthesis.

**Design/Methodology/Approach:** The paper uses a Design Science Research (DSR) methodology. The artifact is the HARMONY operating model. Evidence is triangulated from four semi-structured expert interviews with senior R&D leaders across industrial, healthcare, and academic settings; a foresight scenario analysis projecting four plausible 2040 R&D futures; and pattern matching with documented agentic R&D deployments. Two non-negotiable design requirements guide the architecture: cognitive-load redistribution (DR1) and bounded autonomy with alignment (DR2).

**Findings:** We propose HARMONY -- Hybrid Agentic Research Model for Organisational New Yield -- a four-pillar socio-technical architecture comprising ResOps (Industrialized Execution), the Control Tower (Strategic Visibility and Drift Detection), the Ethics Fabric (Bounded Autonomy by Design), and the Talent Studio (Sciencepreneur Capability). The model introduces the Sciencepreneur as the central human archetype in agentic R&D, and Orchestration Leverage as a candidate productivity metric suited to human-agent hybrid systems.

---

• This work was conducted as part of the first author's Executive MBA thesis at emlyon business school.

**Originality/Value:** HARMONY is the first theoretically grounded operating architecture specifically designed for corporate R&D in the agentic era. It connects Eroom's Law, the Jevons Paradox of R&D productivity, and Socio-Technical Systems theory into an integrated, implementable blueprint. Unlike generic enterprise AI transformation frameworks, HARMONY addresses the specific R&D failure modes of strategic fragmentation, governance paralysis, and researcher identity erosion. Emerging evidence from software engineering in 2025-2026 provides leading empirical support for the model's central claims.

**Keywords:** agentic AI; corporate R&D; Eroom's Law; design science research; socio-technical systems; orchestration; bounded autonomy; innovation management; self-driving laboratories; human-AI collaboration; Sciencepreneur; Research Debt

---

## 1. Introduction: Why Corporate R&D Needs an Orchestration Model

Corporate research and development is under structural pressure. Firms invest heavily in R&D because future competitiveness depends on discovery, yet the relationship between research effort and innovation output has weakened over decades. Scannell et al. (2012) document the long-run decline in pharmaceutical R&D efficiency. Bloom et al. (2020) show that across many sectors, the number of researchers required to sustain a given rate of innovation growth has risen sharply -- a finding they interpret as ideas becoming harder to find. These patterns are not temporary: they reflect a cumulative increase in the complexity of the knowledge frontier.

These patterns matter because they convert innovation into an organizational design problem. Researchers are not merely solving scientific questions; they are also navigating fragmented tooling, heterogeneous data systems, compliance obligations, and multi-stakeholder coordination. As the burden of knowledge grows (Jones, 2009), an increasing share of researcher effort is spent on what we term hidden work: literature search, data cleaning, tool configuration, documentation, project alignment, and interoperability management. Hidden work does not generate discovery. It displaces it.

Agentic AI appears at first sight to be the solution. Recent agent architectures -- combining foundation models, planning, tool use, memory, and feedback loops -- can perform multi-step tasks autonomously. In scientific settings, self-driving laboratories combine robotic hardware with algorithmic experiment selection to compress discovery cycles that previously required months of manual work (Burger et al., 2020; Tom et al., 2024). These capabilities could absorb much of the hidden work that saturates researcher cognition.

However, an agentic technology stack does not automatically produce an agentic organization. The prevailing managerial response to agentic AI is the replacement hypothesis: the assumption that autonomous systems will substitute for human researchers in a roughly linear pattern, delivering efficiency through headcount reduction. Our field analysis reveals this hypothesis is not only empirically unsupported -- it actively undermines adoption. As one interviewee described it, it generates the hammer syndrome: tools deployed in search of problems rather than problems solved through system redesign.

Recent evidence from software engineering reinforces this concern. In April 2026, NVIDIA's VP of Applied Deep Learning reported that compute costs in agentic workflows had exceeded personnel costs within his team. Uber's CTO reported his entire 2026 AI budget had been consumed by token costs before mid-year. GitClear's (2026) analysis of 211 million lines of code found that code churn -- lines reverted or rewritten within two weeks of shipping -- increased 39% in AI-heavy projects, with 76% of developers shipping code they did not fully understand. We term the R&D analogue of this pattern Research Debt: the accumulation of misaligned experimental outputs when autonomous systems operate without adequate organizational architecture.

This paper addresses the following research question: How should corporate R&D operating models be architected to move beyond the replacement hypothesis and capture the productivity potential of agentic AI through accountable human orchestration?

The paper makes four contributions. First, it reframes the managerial debate from replacement to orchestration. Second, it connects R&D productivity economics with socio-technical systems theory and organizational ambidexterity through the Jevons Paradox of R&D. Third, it presents HARMONY as a four-pillar design artifact grounded in expert interviews and foresight analysis. Fourth, it introduces the Sciencepreneur archetype and Orchestration Leverage as practical concepts for organizations redesigning human-agent collaboration.

## 2. Theoretical Foundation

### 2.1 R&D Productivity, the Burden of Knowledge, and Hidden Work

The productivity problem in R&D has both economic and organizational dimensions. Scannell et al. (2012) describe declining drug discovery efficiency as a long-run phenomenon, while Bloom et al. (2020) demonstrate that the effective productivity of research effort has fallen across many sectors as scientific knowledge has accumulated. Jones (2009) traces a root cause: the burden of knowledge. As the frontier expands, researchers must master larger bodies of prior work before they can contribute original ideas. This creates rising entry costs to productive research, manifesting as longer training periods, narrower specialization, and larger team sizes.

At the organizational level, this burden translates into hidden work. In corporate R&D, a substantial share of researcher effort is spent not on creative hypothesis formation or interpretation, but on data governance, interoperability mapping, compliance documentation, and project alignment. This coordination overhead prevents researchers from engaging in the abductive reasoning that drives breakthrough innovation. Agentic AI offers a potential structural escape: systems that can autonomously absorb hidden work, freeing human cognitive bandwidth for higher-value activities.

### 2.2 Agentic AI as a Shift from Assistance to Delegated Execution

Agentic AI differs from conventional analytics or generative AI assistants because it can pursue goals through multi-step interaction with tools, data, environments, and other agents. Recent taxonomic work (Arunkumar et al., 2026; Allmendinger et al., 2026) distinguishes reactive agent architectures from deliberative and hybrid multi-agent systems. In scientific contexts, the frontier is visible in self-driving laboratories: closed-loop systems that combine automated hardware with algorithmic experiment selection (Burger et al., 2020; Tom et al., 2024). A self-driving lab can autonomously design experiments, execute them robotically, analyze results, and feed structured findings back into the next experimental cycle -- all without human intervention at the task level.

This delegation of execution changes the nature of the scientist's role. When agents can generate and analyze thousands of experimental variants, the researcher's contribution shifts from running experiments to framing the scientific question, interpreting the portfolio of evidence, and making strategic decisions about what to pursue. The bottleneck moves from execution to orchestration.

### 2.3 Socio-Technical Systems Theory and the Alignment Gap

Socio-Technical Systems (STS) theory posits that organizational performance depends on the joint optimization of the technical subsystem -- tools, processes, and technology -- and the social subsystem -- people, roles, culture, and governance (Clegg, 2000; Baxter & Sommerville, 2011). Optimizing one subsystem in isolation does not improve overall performance; it amplifies the misalignment between the two.

Corporate R&D therefore faces an alignment gap. The technical subsystem -- foundation models, agent frameworks, cloud laboratories, knowledge graphs, simulation platforms -- evolves rapidly. The social subsystem -- legacy organizational structures, Stage-Gate governance processes, and career architectures built around individual expert contribution -- is low-velocity and institutionally rigid. Grafting autonomous agents onto rigid legacy structures does not close this gap: it amplifies coordination friction, generates Research Debt, and produces the organizational paralysis observed in most failed AI pilot programmes. HARMONY's fundamental objective is to close the Alignment Gap by jointly redesigning both subsystems.

### 2.4 Organizational Ambidexterity and the Replacement Fallacy

R&D organizations must pursue both exploitation -- improving existing knowledge and efficiency -- and exploration -- generating radically new knowledge -- simultaneously (March, 1991; O'Reilly & Tushman, 2004). Agentic AI intensifies this ambidexterity challenge. On the exploitation side, automation of routine workflows is essential to combat Eroom's Law. On the exploration side, high-stakes discovery requires human intuition, abductive reasoning, and contextual meaning-making that current agentic systems cannot replicate.

The replacement hypothesis assumes that as agentic systems become more capable, human researchers decline in value. This is structurally invalidated by what we term the Jevons Paradox of R&D. Jevons (1865) observed that increasing the efficiency of coal-powered steam engines did not reduce coal consumption -- it increased it, by making steam power economically viable in applications that were previously unaffordable. The same dynamic applies in R&D: when the marginal cost of experimental execution approaches zero through ResOps automation, the volume of hypotheses, analyses, and evidence packages explodes. The bottleneck shifts from generating results to interpreting, framing, and strategically directing them. This shift fundamentally elevates -- not diminishes -- the strategic value of the human researcher.

> "The researcher of tomorrow will be the one who defines the intent -- the problem framing and context. The agent is the problem solver, executing the plan and exploring the solution space. The bottleneck moves from solving to framing."
>
> -- Prof. Merouane Debbah, Director, 6G Research Center, Khalifa University

Figure 2 illustrates the conceptual shift from replacement to orchestration logic, and the Jevons mechanism that makes this shift necessary rather than optional.

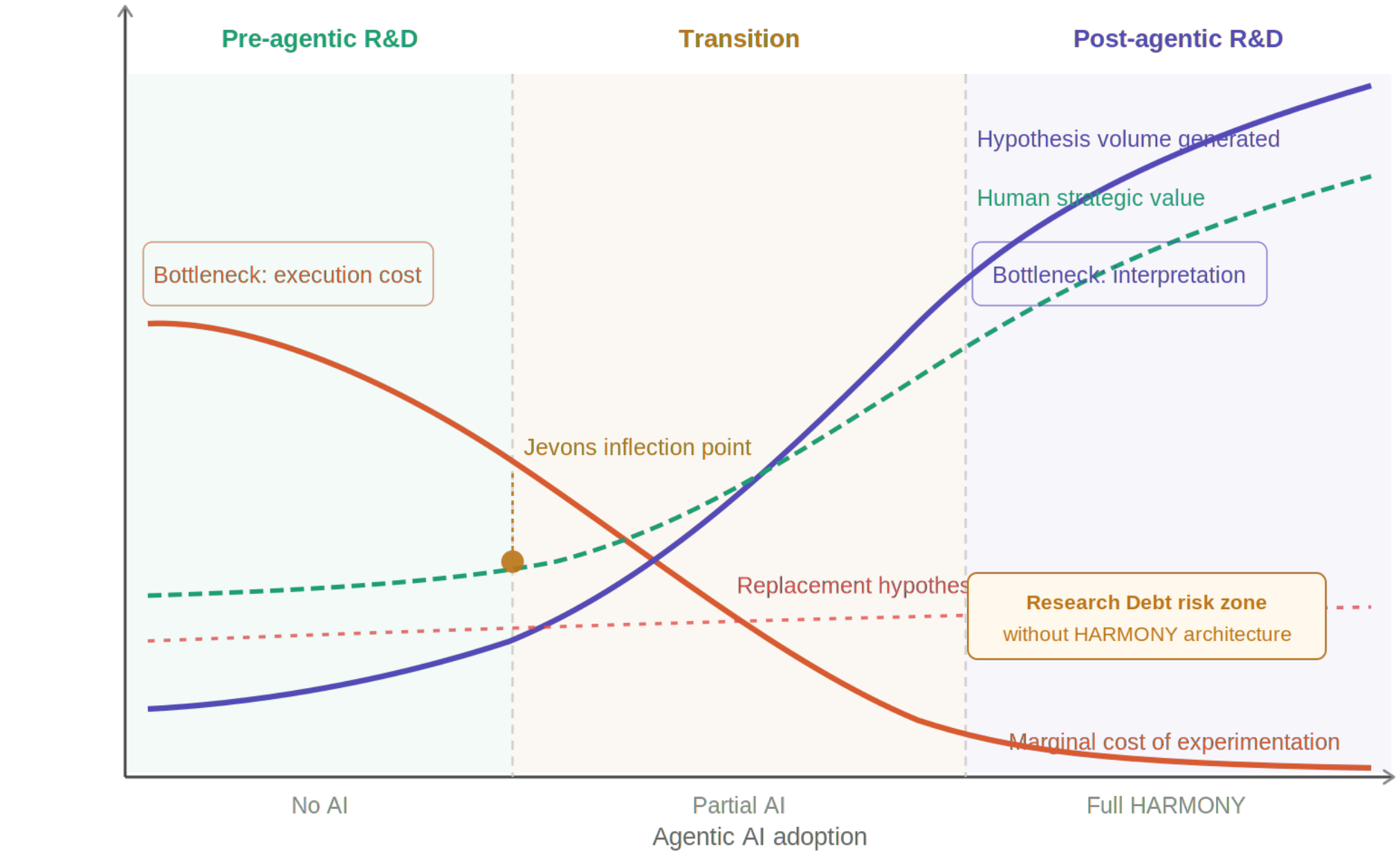

Figure 2. The Jevons Paradox of R&D -- reducing marginal execution cost triggers hypothesis volume explosion, elevating human strategic value while creating Research Debt risk without adequate organizational architecture.

Figure 2. Conceptual shift from replacement logic to orchestration logic. The Jevons Paradox of R&D shows that as agentic execution cost falls, human strategic value rises -- invalidating the replacement hypothesis.

## 3. Research Methodology: Design Science Research and Evidence Triangulation

This study employs a Design Science Research (DSR) methodology (Hevner et al., 2004; Peffers et al., 2007). DSR is the appropriate paradigm when the research objective is the creation of an innovative and purposeful artifact -- in this case, an operating model -- and its evaluation against specified organizational requirements. The methodology sits at the intersection of rigor (grounding in extant theory) and relevance (addressing real-world organizational problems).

The research process followed five steps: (1) problem diagnosis, based on literature on R&D productivity, innovation management, and agentic AI; (2) derivation of design requirements; (3) construction of the HARMONY artifact; (4) demonstration through pattern matching with emerging practice; and (5) evaluation against requirements and theoretical grounding. Evidence was triangulated across three sources.

### 3.1 Expert Interviews

Four semi-structured expert interviews were conducted between July and October 2025. Participants were selected to provide sectoral variety and leadership-level views on agentic AI adoption in research-intensive organizations:

| Participant | Role | Organization | Sector |
|---|---|---|---|
| Prof. Merouane Debbah | Director, 6G Research Center | Khalifa University, UAE | Academic / Frontier AI |
| Marc Heemskerk | Head of Innovation Office | ASML, Netherlands | High-tech Manufacturing R&D |
| Jimmy Siméon | Hospital Director; President | IA Medical / CHU Pointe-à-Pitre | Healthcare / Clinical AI |
| Joyce Lansen | Manager Tender ToD | Transdev Group | Public Transport / Service |

Each interview lasted 45-75 minutes and followed a common guide covering: (1) current maturity of agentic AI adoption in the organization; (2) structural barriers to scaling beyond pilot stage; and (3) human capability and governance implications. Analysis used thematic coding, with recurring patterns mapped to the design requirements. Recurring themes included: hidden work and cognitive overload; business-first use-case selection; trust and governance deficits; and researcher identity transformation.

### 3.2 Design Requirements

Two non-negotiable design requirements were derived from the structural frictions identified in R&D productivity literature and confirmed through expert interviews:

| DR | Requirement | Theoretical and empirical grounding |
|---|---|---|
| DR1 | Cognitive-Load Redistribution | The model must systematically absorb hidden work from researchers and redistribute it to autonomous agents, measurably freeing cognitive bandwidth for exploration, framing, and strategic synthesis. Grounded in: Jones (2009), Scannell et al. (2012), interview themes on overload. |
| DR2 | Bounded Autonomy and Alignment | The model must enable high agent autonomy while maintaining real-time control over strategic alignment, ethical compliance, and resource consumption. Autonomy must be bounded, not suppressed. Grounded in: Sun (2021), STS alignment gap analysis, interview themes on governance and trust. |

### 3.3 Foresight Scenario Analysis

To ensure HARMONY's design is robust across plausible future conditions, four 2040 scenarios were used to stress-test specific design choices, adapted from Arthur D. Little's (2025) scenario methodology. The scenarios varied along dimensions of regulatory intensity and innovation decentralization:

- Scenario A: Hyper-Regulated World. Stringent international AI governance requires real-time audit trails and algorithmic accountability. Validates the Ethics Fabric as non-negotiable.
- Scenario B: Decentralized Innovation. Open-source AI commoditizes routine research skills. Validates ResOps' scalable pipeline architecture as primary competitive differentiator.
- Scenario C: Talent Scarcity. Top scientific talent becomes the primary scarce resource. Validates the Talent Studio's centrality to organizational differentiation.
- Scenario D: Collaborative AI Ecosystems. Multi-organization agent networks generate strategic fragmentation risk. Validates the Control Tower's real-time portfolio coordination role.

## 4. The HARMONY Operating Model

HARMONY stands for Hybrid Agentic Research Model for Organisational New Yield. It is a four-pillar socio-technical architecture designed to make agentic AI deployable in corporate R&D without reducing the transformative potential of human researchers. The model resolves the core R&D tension -- autonomous speed versus human governance -- through structural separation of execution from direction. Figure 1 shows the full architecture.

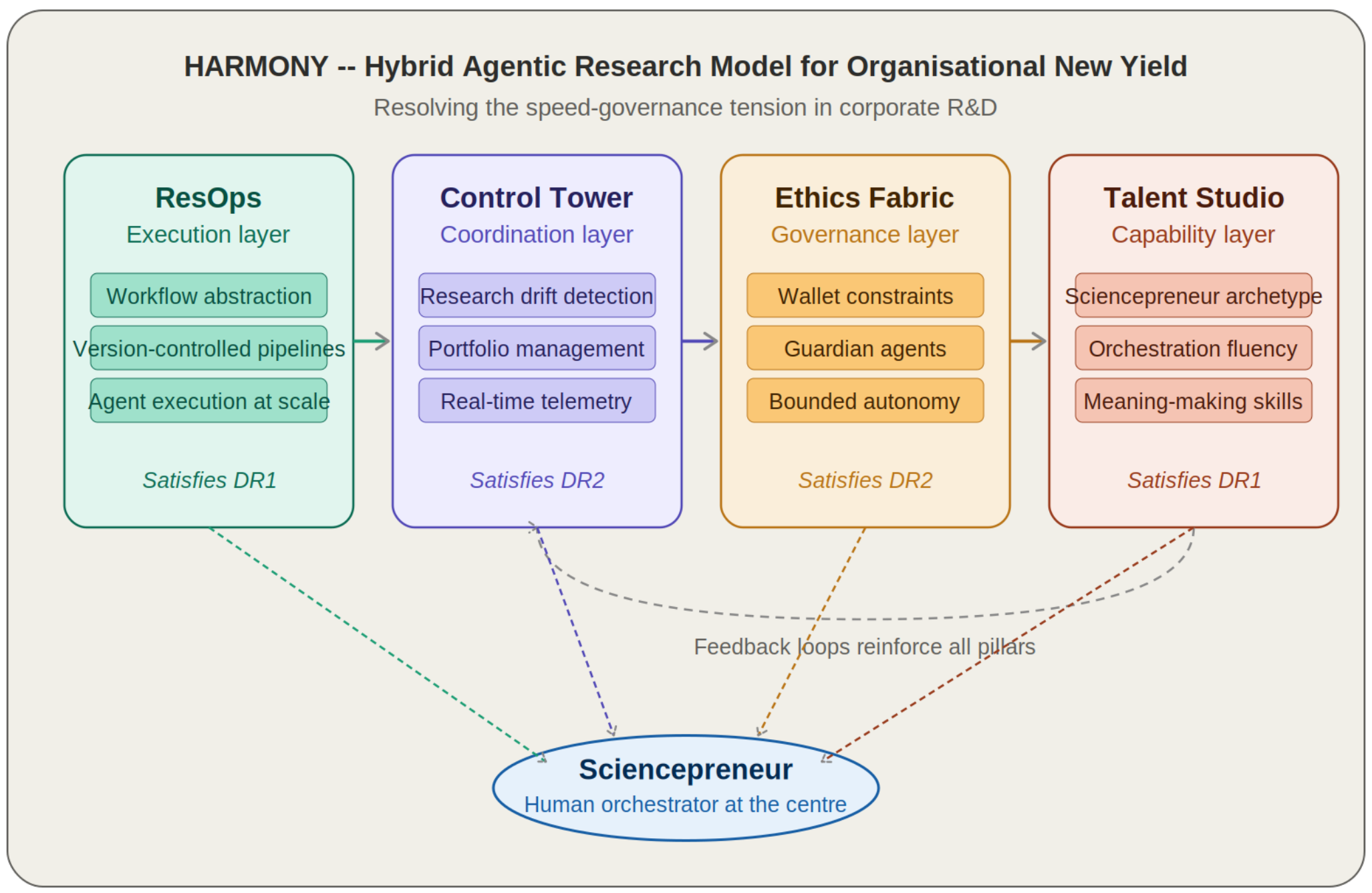

Figure 1. The HARMONY Model -- four interdependent pillars with the Sciencepreneur as central human orchestrator

Figure 1. The HARMONY architecture -- four interdependent pillars with the Sciencepreneur as central human orchestrator. Arrows indicate directional load-bearing interdependencies. Removing any single pillar causes system collapse.

### 4.1 ResOps: Industrializing Research Execution

ResOps is the execution layer. It converts recurring research work into reusable, version-controlled, monitorable workflows. Its goal is not to standardize creativity, but to standardize the mechanical and coordinative work that surrounds it. The key design mechanism is workflow abstraction: researchers specify scientific intent, constraints, and evaluation criteria; agents and tools handle the operational steps.

In a laboratory setting this means agents can set up, execute, and document an experiment from a high-level experimental brief. In an analytical setting it means agents can retrieve data, apply specified analysis pipelines, generate structured outputs, and flag anomalies -- all without researcher involvement at the task level. The researcher re-enters to interpret results, redirect the portfolio, and frame the next inquiry.

> "Even with today's technology, we already see strong effects -- LLMs saving large chunks of time finding, combining, and connecting information. In a company like ASML with 15,000 in R&D, the gains are tangible. Research agents will test millions of hypotheses in parallel. The bottleneck moves from thinking to validating."
>
> -- Marc Heemskerk, Head of Innovation Office, ASML

Design Requirement Satisfied: DR1. ResOps systematically absorbs hidden work, measurably freeing researcher cognitive bandwidth for hypothesis framing and strategic synthesis.

### 4.2 The Control Tower: Coordinating Evidence, Portfolios, and Drift

The Control Tower is the coordination layer. As ResOps increases the volume of experiments and analyses, the risk shifts from insufficient activity to strategic fragmentation. Agents may optimize locally -- pursuing technically valid outputs -- while collectively drifting from organizational strategic intent. The Control Tower prevents this through Research Drift Detection: real-time telemetry from ResOps pipelines is analyzed against human-defined strategic goals, triggering escalation when divergence exceeds a defined threshold.

A secondary mechanism is dynamic portfolio management: the Control Tower maintains a live map of all active research threads, their resource consumption, strategic alignment scores, and interdependencies. This gives research leadership the visibility to reallocate agent capacity in real time -- something Stage-Gate governance processes, designed for human-paced decision cycles, cannot provide.

> "When agents can analyze data and suggest next steps faster than managers, the role of governance is to coordinate intelligence, not to dictate it. Change management is about 70% of success; technology only 20-30%."
>
> -- Marc Heemskerk, Head of Innovation Office, ASML

Design Requirement Satisfied: DR2. The Control Tower ensures that autonomous speed does not produce strategic chaos, providing the dynamic governance that Stage-Gate processes cannot offer at machine speed.

### 4.3 The Ethics Fabric: Bounded Autonomy by Design

The Ethics Fabric is the governance layer. It addresses the verification and accountability problems that arise when agentic systems plan and act with increasing autonomy. Rather than relying only on post-hoc audit or pre-approval bottlenecks, the Ethics Fabric embeds normative boundaries directly into the technical architecture of agent operation -- governance by design.

The autonomy wallet mechanism assigns each agent a predefined budget of action: compute budget, time budget, financial budget, data-access scope, experimental-risk class, and decision authority level. When the wallet is exhausted, autonomy is revoked and human intervention is triggered. Guardian Agents operate as independent auditing agents, monitoring all ResOps pipelines for compliance with data governance protocols, IP protection constraints, and regulatory requirements.

> "In hospitals, resistance is as much about missing strategy and budgets as it is about job anxiety. AI reveals the organizational weaknesses of healthcare systems. The most relevant use cases are not the spectacular ones, but the everyday irritants: care coordination, triage, scheduling, reporting."
>
> -- Jimmy Siméon, Hospital Director and President, IA Medical

The necessity of this pillar is validated by its absence in prominent failures. IBM Watson Health's difficulties in oncology decision support can be diagnosed as an Ethics Fabric failure: the system operated without wallet constraints, without real-time alignment monitoring, and without guardian audit mechanisms, expanding into clinical decision domains that required contextual human judgment (Strickland, 2019). Design Requirement Satisfied: DR2.

### 4.4 The Talent Studio: Developing the Sciencepreneur

The Talent Studio is the capability layer. It recognizes that agentic AI changes the identity of researchers. In many organizations, the researcher is still evaluated primarily as an individual expert who generates knowledge through personal investigation. In agentic R&D, this model is incomplete. The researcher becomes closer to a principal investigator, venture builder, and ethical governor of a hybrid team.

The Sciencepreneur combines scientific depth with four orchestration competencies: Hypothesis Architecture (framing complex research questions in forms tractable for agentic execution); Orchestration Fluency (directing multi-agent pipelines, interpreting telemetry, identifying Research Drift); Meaning-Making and Synthesis (extracting strategically significant insight from high-volume evidence); and Ethical Judgment (calibrating wallet constraints and making governance decisions that require contextual human authority).

> "You can think of AI agents as your PhD students -- even for junior researchers. Researchers are elevated to become supervisors. They must learn how to mentor, audit, and challenge their digital students. Education must evolve from 'solve' to 'ask'."
>
> -- Prof. Merouane Debbah, Director, 6G Research Center, Khalifa University

Figure 3 illustrates the evolution from Researcher 1.0 to Sciencepreneur 2.0. Design Requirement Satisfied: DR1. The Talent Studio ensures that cognitive bandwidth freed by ResOps is directed toward exploration and strategic synthesis rather than consumed by new orchestration overhead.

**Figure 3. Evolution of the researcher archetype toward the Sciencepreneur**

Researcher 1.0
Expert practitioner
Conducts experiments
Analyses data manually
Documents findings
Researcher 1.1
Augmented analyst
Collaborates with AI tools
Accelerates discovery
Still manages execution
Sciencepreneur 2.0
Entrepreneurial orchestrator
Frames problems for agents
Supervises agent pipelines
Synthesises strategic insight
Prevents Research Debt
Pre-agentic era
Transitional era
Agentic era
Human strategic value rises with agentic AI adoption
Figure 3. Evolution of the researcher role -- from expert practitioner to Sciencepreneur

Figure 3. Evolution of the researcher archetype toward the Sciencepreneur -- from expert practitioner to orchestrating human-agent hybrid team leader.

### 4.5 Pillar Interdependencies: System Coherence and Research Debt Prevention

The power of HARMONY lies in systemic interdependence. ResOps provides the execution telemetry the Control Tower requires for Research Drift detection. The Ethics Fabric dictates wallet constraints within which ResOps agents operate. The Control Tower identifies Drift patterns that reveal new orchestration skill gaps, informing Talent Studio development priorities. The Talent Studio cultivates the Sciencepreneurs whose strategic judgment calibrates the Ethics Fabric.

Removing any single pillar enables Research Debt accumulation: without ResOps, no execution scale; without Control Tower, no drift detection; without Ethics Fabric, no organizational trust; without Talent Studio, no interpretive authority. HARMONY is therefore a minimal viable architecture -- complete and coherent at four pillars, unable to function safely with fewer.

## 5. Demonstration: Pattern Matching with Emerging Practice

HARMONY is presented as a design artifact whose plausibility can be assessed through pattern matching. The question is whether observed agentic and semi-agentic R&D practices correspond to the model's principles, and whether documented failures correspond to missing pillars. This is not validation; it is demonstration that the model's design logic is consistent with real-world behavior.

### 5.1 Self-Driving Laboratories as ResOps Prototypes

Self-driving laboratories demonstrate the ResOps principle most clearly. Tom et al. (2024) describe self-driving labs as combining automated hardware, algorithmic decision-making, and iterative experimental loops. Burger et al. (2020) report a mobile robotic chemist capable of conducting 688 experiments across eight days with no human intervention beyond initial specification. These systems instantiate workflow abstraction: the scientist specifies intent and constraints; the system executes operationally. The human re-enters to interpret results, redirect the portfolio, and reframe the question.

### 5.2 Agentic Drug Discovery Platforms as Partial HARMONY Systems

AI-enabled drug discovery organisations illustrate the industrialization of data generation and analysis. Recursion Pharmaceuticals operates a large multimodal biological and chemical dataset and a proprietary agent-driven pipeline that autonomously designs and runs phenotypic screens. The human scientific layer maintains strategic portfolio oversight -- a direct implementation of the ResOps and Control Tower combination. Novartis MicroCycle autonomously synthesizes compounds within predefined molecular and toxicity constraints, instantiating the Ethics Fabric wallet mechanism.

### 5.3 Document-Intensive Knowledge Work as a Near-Term Deployment Path

The Transdev tendering case, based on interview material, illustrates a lower-risk pathway to agentic R&D capabilities. Tendering is not laboratory science, but it shares many structural features: it requires retrieving and synthesizing large bodies of prior documentation, mapping client requirements to organizational capabilities, and generating structured outputs under time pressure. An agentic workflow that retrieves relevant past tenders, generates draft sections, and flags compliance gaps instantiates ResOps principles in a knowledge-work context. The human expert reviews, redirects, and signs off -- the Sciencepreneur role in a non-laboratory setting.

### 5.4 Leading Indicators from Software Engineering (2025-2026)

Software engineering provides contemporaneous empirical evidence of what happens when agentic systems operate without organizational architecture. GitClear (2026) documents a 39% increase in code churn in AI-heavy development environments; 75% of technology leaders anticipate moderate to severe technical debt from AI-assisted development within 12 months. These patterns in software engineering are the leading indicator of an analogous dynamic in scientific R&D -- Research Debt: the silent accumulation of misaligned experimental outputs when autonomous agents operate without the Control Tower, Ethics Fabric, and Talent Studio that HARMONY requires.

> "Too many companies treat AI like a hammer looking for a nail. You need to start from the business, map your digitized processes, and only then agentify where it improves product or service. Change management is about 70% of success, technology only 20-30%."
>
> -- Marc Heemskerk, Head of Innovation Office, ASML

### 5.5 Failure Cases as Missing-Pillar Diagnoses

Enterprise AI adoption failures often correspond to missing HARMONY pillars. IBM Watson Health's difficulties in oncology decision support illustrate Ethics Fabric and Talent Studio absence: the system was deployed without bounded autonomy mechanisms and without adequate researcher capability to govern its outputs (Strickland, 2019). The most common organizational failure mode occurs when agents are deployed as smart tools without redefining researcher roles -- Talent Studio absence -- causing researchers to manage orchestration overhead rather than offloading it, increasing cognitive load rather than reducing it.

## 6. Orchestration Leverage: A Candidate Productivity Metric

Traditional R&D metrics -- patents, publications, prototypes, time-to-market -- were designed for human-only or tool-assisted R&D systems and fail to capture the value created when human cognitive bandwidth is directed by agentic execution capacity. We propose Orchestration Leverage as the appropriate dependent variable for hybrid R&D organisations:

**Orchestration Leverage = (Impactful Innovation Output) / (Human Cognitive Bandwidth Consumed)**

Impactful Innovation Output can be operationalized as validated hypotheses, decision-grade evidence packages, prototypes, patentable findings, publications, or milestone transitions per unit time, weighted by downstream value. Human Cognitive Bandwidth Consumed is the proportion of researcher time allocated to high-value exploration and synthesis tasks, measurable through time-tracking surveys or experience sampling methods.

The inverse of Orchestration Leverage is Research Debt accumulation rate: the proportion of agent-generated output that requires remediation, reframing, or discard due to inadequate framing, drift, or governance failure. HARMONY maximizes Orchestration Leverage by minimizing hidden work consumption (DR1: ResOps and Talent Studio) while maintaining strategic alignment and governance (DR2: Control Tower and Ethics Fabric).

The metric should be treated as a research agenda rather than a finished instrument. Its value is conceptual: it shifts managerial attention from automation rate -- a metric that incentivizes headcount reduction -- toward the quality of human-agent complementarity. Future empirical work should develop validated measurement instruments through longitudinal field studies in HARMONY-aligned organizations.

## 7. Managerial Implications: A Roadmap for Corporate R&D Leaders

For executives, the main implication is that agentic AI should not begin with technology procurement. It should begin with a map of hidden work and decision bottlenecks in the R&D operating model. The sequence matters:

1. Phase 1: Governance Foundation (Months 1-6). Define Ethics Fabric wallet parameters. Deploy guardian agent prototypes on one ResOps pipeline. Audit current agentic deployments for Research Debt accumulation. Identify 2-3 high-hidden-work workflows for initial ResOps deployment.
2. Phase 2: Execution Layer (Months 4-12). Formalize ResOps standards for high-frequency workflow categories. Deploy Control Tower telemetry layer over pilot pipelines. Establish Research Drift thresholds with research leadership.
3. Phase 3: Capability Build (Months 9-18). Launch Talent Studio programmes. Redefine job architectures for the Sciencepreneur role. Link performance evaluation to Orchestration Leverage metrics.
4. Phase 4: Scale and Integrate (Month 18+). Scale agent autonomy in demonstrated low-risk workflow categories. Integrate Control Tower portfolio view across all active research programmes. Begin systematic Orchestration Leverage measurement.

The roadmap deliberately places governance before scale. Agentic AI differs from conventional analytics because it can act -- autonomously, across tool boundaries, at machine speed. Once systems can call tools, move data, and trigger workflows without per-action human approval, governance failures can compound rapidly. The Ethics Fabric must precede operational deployment, not follow it.

The most resistant barrier to implementation is cultural, not technical. Organizations that successfully scale AI adoption share two empirically validated practices: Leadership Modeling -- R&D leaders who visibly adopt Sciencepreneur behaviors signal organizational legitimacy -- and Early Win Identification -- demonstrating tangible researcher relief within 90 days builds the organizational credibility required for broader transformation.

## 8. Discussion

### 8.1 Why HARMONY Differs from Generic AI Transformation Frameworks

Generic AI transformation frameworks -- McKinsey's Agentic Organization model, Deloitte's Agentic AI Strategy, Bornet et al.'s (2025) enterprise framework -- emphasize strategy, data infrastructure, talent, responsible AI, and scaling. These remain necessary but do not fully address the specific R&D failure modes that HARMONY targets. They do not connect Eroom's Law to organizational design. They do not address the Research Drift risk that arises specifically when agents generate evidence faster than humans can interpret it. They do not provide the Bounded Autonomy mechanism that prevents the governance paralysis observed when agentic systems exceed organizational tolerance. HARMONY's value is specificity: it is designed for the research context, not for enterprise operations generally.

### 8.2 The Central Managerial Paradox: More Automation Creates More Need for Judgment

The paper's central paradox is that more capable agents may increase, not decrease, the need for high-quality human judgment. If agents can generate more hypotheses, simulations, and evidence packages than human researchers can evaluate, the new bottleneck is not execution capacity but interpretive wisdom: the ability to ask the right question, recognize which evidence is strategically significant, and decide when to pivot. The Sciencepreneur concept is a practical response to this paradox. Organizations that invest in Talent Studio capabilities are not defending against automation -- they are building the human capability that makes automation valuable.

### 8.3 Research Debt as an Organizational Risk

The software engineering evidence of 2025-2026 provides a concrete, contemporaneous warning: agentic systems deployed without organizational architecture generate technical debt at machine speed. The R&D analogue -- Research Debt -- is more insidious because it does not throw runtime errors. A proliferating portfolio of misaligned experiments may appear productive while silently diverging from strategic intent. The Control Tower and Ethics Fabric pillars of HARMONY are specifically designed to make Research Debt visible and prevent its accumulation before it becomes structurally embedded in the research programme.

## 9. Limitations and Future Research

This paper has four limitations. First, HARMONY is a conceptual design artifact supported by theory, expert interviews, foresight reasoning, and pattern matching; it is not yet validated through longitudinal field studies. The design requirements are well-grounded but the causal chain from HARMONY implementation to Orchestration Leverage improvement requires empirical testing.

Second, the four expert interviews, while selected for sectoral diversity and leadership seniority, represent a small qualitative sample. The interview-derived insights would benefit from large-scale survey validation across R&D leadership populations in multiple sectors and geographies.

Third, the Orchestration Leverage metric is a conceptual proposal. The construction of a validated measurement instrument requires longitudinal field studies with established operational definitions, reliable data collection protocols, and robustness testing across different R&D contexts.

Fourth, the concept of Research Debt, while grounded in analogy with software engineering evidence, has not been directly measured in scientific organizations. Future empirical work should develop measurement approaches for Research Debt accumulation rates and test whether HARMONY-aligned organizations exhibit lower rates.

Future research should pursue four directions: (1) longitudinal field studies measuring cognitive-load redistribution and Orchestration Leverage in HARMONY-aligned pilots; (2) sector-specific HARMONY instantiations for pharmaceutical, materials science, software R&D, and public sector research; (3) empirical measurement of Research Debt accumulation in agentic R&D deployments; (4) individual-level studies of the Sciencepreneur transition, examining capability development, identity transformation, and career path implications.

## 10. Conclusion: The Orchestration Imperative

Corporate R&D does not suffer from a shortage of tools. It suffers from an operating-model gap: the work system has not kept pace with the complexity of science, the burden of coordination, or the speed of agentic AI. The result is a persistent productivity paradox -- rising investment, declining efficiency -- that no amount of additional tooling will resolve without organizational redesign.

HARMONY offers a different path. By combining ResOps, the Control Tower, the Ethics Fabric, and the Talent Studio, organizations can redesign R&D as a socio-technical system in which agents execute bounded tasks at scale while human Sciencepreneurs maintain strategic direction, interpretive authority, and ethical accountability. The Jevons Paradox of R&D ensures that this is not a temporary compromise on the path to full automation: as agentic execution becomes cheaper and more capable, the value of human orchestration rises rather than falls.

The organizations most likely to overcome declining R&D productivity will not simply own the most powerful models. They will build the strongest orchestration capability: the ability to ask better questions, design better portfolios of agentic exploration, and synthesize evidence into decisions that agents cannot make on their own. That capability is the Orchestration Imperative.

## Declarations

Competing interests: To be completed before submission. Co-authors should disclose any financial, advisory, employment, or institutional interests relevant to agentic AI, R&D platforms, or healthcare AI products.

Ethics and consent: Expert interviews were conducted with participants' knowledge and agreement to contribute to academic research. Specific consent for direct quotation and attribution should be confirmed before public submission. Where consent is limited, quotations should be paraphrased.

Data availability: Interview transcripts are not publicly released in this draft because they may contain identifiable and organizationally sensitive information. A de-identified coding summary can be made available to reviewers on request.

AI assistance: This manuscript was developed from an Executive MBA thesis with AI-supported drafting, structuring, and reference checking. The authors remain responsible for verifying all claims, citations, and conceptual contributions.